\documentclass[useAMS]{mn2e}
\usepackage{times}
\usepackage{epsfig}

\newcommand{\mdot}{\mbox{$\dot{M}$}}
\newcommand{\vinf}{\mbox{$v_{\infty}$}}

\def \etal   {\hbox{et~al.\/}}
\def \kms    {km~s$^{-1}$}
\def \msunyr   {{$M_{\odot}$~yr$^{-1}$}}
\def\lesssim{\mathrel{\hbox{\rlap{\hbox{\lower4pt\hbox{$\sim$}}}\hbox{$<$}}}}
\def\gtrsim{\mathrel{\hbox{\rlap{\hbox{\lower4pt\hbox{$\sim$}}}\hbox{$>$}}}}

\title[Forbidden Lines from Colliding Winds]
{Modeling Forbidden Line Emission Profiles from 
Colliding Wind Binaries}

\author[Ignace, Bessey, \& Price ]
{
R.~Ignace, R.~Bessey\footnote{Currently at Georgia Tech}, C.~S.~Price\\
       Department of Physics \& Astronomy,
       East Tennessee State University,
       Box 70652,
       Johnson City, TN 37614
       USA
}

\begin{document}
\maketitle


\begin{abstract}

This paper presents calculations for forbidden emission line profile
shapes arising from colliding wind binaries.  The main application
is for systems involving a Wolf-Rayet (WR) star and an OB star
companion.  The WR wind is assumed to dominate the forbidden line
emission.  The colliding wind interaction is treated as an archimedean
spiral with an inner boundary.  Under the assumptions of the model,
the major findings are as follows.  (a) The redistribution of the
WR~wind as a result of the wind collision is not flux conservative
but typically produces an excess of line emission; however, this
excess is modest at around the 10\% level.  (b) Deviations from a
flat-top profile shape for a spherical wind are greatest for viewing
inclinations that are more nearly face-on to the orbital plane.  At
intermediate viewing inclinations, profiles display only mild
deviations from a flat-top shape.  (c) The profile shape can be
used to constrain the colliding wind bow shock opening angle.  (d)
Structure in the line profile tends to be suppressed in binaries
of shorter periods.  (e) Obtaining data for multiple forbidden lines
is important since different lines probe different characteristic
radial scales.  Our models are discussed in relation to {\em Infrared
Space Observatory} data for WR~147 and $\gamma$~Vel (WR~11).  The
lines for WR~147 are probably not accurate enough to draw firm
conclusions.  For $\gamma$~Vel, individual line morphologies are
broadly reproducible but not simultaneously so for the claimed wind
and orbital parameters.  Overall, the effort demonstrates how lines
that are sensitive to the large-scale wind can help to deduce binary
system properties and provide new tests of numerical simulations.

\end{abstract}


\section{Introduction}

Massive star colliding wind systems continue to be an area of intense
study.  The colliding wind interaction region (CWIR) is intrinsically
interesting as a problem of hydrodynamic flow (Girard \& Willson
1987; Shore \& Brown 1988; Stevens, Blondin,
\& Pollock 1992; Usov 1992; Canto, Raga, \& Wilkin 1996; Gayley,
Owocki, \& Cranmer 1997).  The CWIRs lead to a number of interesting
observable effects as well, such as the X-ray emissions (e.g.,
Stevens \etal\ 1996; Maeda \etal\ 1999; Zhekov \& Skinner 2000;
Pittard \etal\ 2002; Henley, Stevens, \& Pittard 2003), ultraviolet
(UV), optical, and infrared (IR) emission line profile effects (e.g.,
Stevens 1993; L\"{u}hrs 1997; Stevens \& Howarth 1999; Hill \etal\ 2000;
Hill, Moffat, \& St-Louis 2002), dust emission
(e.g., Tuthill, Monnier, \& Danchi 1999; Monnier, Tuthill, \& Danchi 1999;
Tuthill \etal\ 2006; Tuthill \etal\ 2008), and polarization (e.g., Brown,
McLean, \& Emslie 1978; Drissen \etal\ 1986; Villar-Sbaffi \etal\ 2003,
2005, 2006).  Of chief interest between relating the observables to the
hydrodynamical predictions are the determinations of the mass-loss rates
for the two stars, their orbital properties, and the opening angle of
the colliding wind bow shock.  Obtaining the orbital solution implies
deriving the masses of the two stars, which is important for understanding
massive star evolution.  As is typical with unresolved binary systems,
the viewing inclination $i$ creates a challenge to obtaining the full
orbital solution and stellar masses.

In the study of CWIRs of massive binaries, the interpretation of
optically thin lines remains prominent, such as X-ray lines that
encode information about opening angle and viewing inclination.
One area that has been largely ignored is the influence of the CWIR
on the {\em shape} of forbidden emission line profiles, especially
those in the IR.  The {\em Infrared Space Observatory} (`{\em ISO}')
and the {\em Spitzer Space Telescope} ('{\em Spitzer}') have provided
a large quantity of high quality IR spectra for massive stars.
Especially in the case of the evolved Wolf-Rayet (WR) stars, the
IR spectra are rich in emission lines and in particular forbidden
lines.  These forbidden lines have been useful for deriving or
constraining gaseous abundances that are relevant for testing models
of massive star evolution (e.g., Barlow, Roche, \& Aitken 1988;
Willis \etal\ 1997; Dessart \etal\ 2000; Morris \etal\ 2000; Morris,
Crowther, \& Houck 2004 ; Smith \& Houck 2005; Ignace \etal\ 2007).

An interesting property of these forbidden lines is that they form
at large radius in the wind of a WR star.  Consequently, the line
formation occurs well beyond the wind acceleration thus sampling
the flow at the wind terminal speed $v_\infty$.  Being optically
thin these forbidden lines should be flat-topped in shape, and so
they are excellent for measuring $v_\infty$ values (Barlow \etal\ 1988).

But the flat-topped profile morphology only holds if the wind is
spherically symmetric.  If a WR star is part of a binary, then even
if the WR wind is intrinsically spherically symmetric, the CWIR ensures
that it will not remain so.  Equally interesting is the fact that
forbidden lines can form over tens of thousands of WR stellar radii
(e.g., Ignace \etal\ 2001), corresponding to scales of several AU and
comparable to binary separations in massive star binaries.  As a
result, forbidden emission profiles should not be flat-topped in
general in these systems, and their deviation from flat-top -- like
the case for other CWIR emission line diagnostics -- can be used
to constrain the properties of the CWIR geometry.  

In this paper we calculate a range of forbidden line profile shapes
as a function of wind and orbit parameters.  Section 2 presents a
brief overview of forbidden line formation, a derivation of emission
profiles in a spherical wind, and then in a colliding wind system.
Then Section 3 describes applications to two systems:  WR~147 and
$\gamma$~Vel.  Concluding remarks are given in Section 4.

\section{Forbidden Line Formation in Winds}

Our presentation of the theory of forbidden line emission generally
follows Osterbrock (1989) regarding the atomic physics, Barlow
\etal\ (1988) for application to winds, and Ignace \&
Brimeyer (2006) for notation.  The following sections describe (a)
the two-level approximation for forbidden lines, (b) the solution
for the forbidden line emission in a spherical wind, (c) the adopted
model used to describe a CWIR between a WR~star and an OB star, and
(d) adjustments to the forbidden line profile calculation arising from the
CWIR.

\subsection{The Atomic Physics}

For simplicity the two-level atom approximation is adopted for a
fine structure transition of an ion species.  The lower level will
be `1' and the upper level `2'.  It is common to introduce a critical
density $n_{\rm c}$ that signifies the transition from the higher
density medium where de-excitations are dominated by collisions
versus the lower density zone where spontaneous decay dominates.
Since excitation is assumed to derive from collisions {\em only},
the emissivity transitions from a function that is linear in density
in the collisional regime to one that is quadratic in density in
the decay regime.  The critical density is defined as

\begin{equation}
n_{\rm c} = \frac{A_{21}}{q_{21}}.
	\label{eq:crit}
\end{equation}

\noindent where $A_{21}$ [s$^{-1}$] is the Einstein A-value for the
transition, and $q_{21}$ is the downward volume collisional
de-excitation rate [cm$^3$ s$^{-1}$].  The volume emissivity $j$
[erg s$^{-1}$ cm$^{-3}$ sr$^{-1}$] is

\begin{equation}
j = \frac{1}{4\pi}\,h \, \nu_{21} \, A_{21}\, n_2.
	\label{eq:emiss}
\end{equation}

\noindent In the two level atom, the entire density of elemental
species $E$ in ion stage $i$ is given by $n_{\rm E,i} = n_1 + n_2$.
One can derive the ratio $n_2/n_1$ from equilibrium conditions and
use that to solve for the population of the upper level: (Barlow
\etal\ 1988):

\begin{equation}
n_2 = \frac{n_{\rm i,E}\,n_{\rm e}\,q_{12}}{n_{\rm e}\,q_{12}
	+n_{\rm e}\,q_{21}+A_{21}}.
	\label{eq:n2}
\end{equation}

\noindent where

\begin{equation}
\varpi \equiv \frac{q_{21}}{q_{12}} = \frac{g_1}{g_2}\,e^\beta,
\end{equation}

\noindent with 

\begin{equation}
\beta = h\nu_{21}/kT_{\rm e}, 
\end{equation}

\noindent for $\nu_{21}$ the frequency of the line transition, $T_{\rm e}$
the electron temperature, and $g_J$ the statistical weight of the level,
with $g_J=2J+1$.  The downward collisional volume rate is given by 

\begin{equation}
q_{21} = \frac{8.629\times10^{-6}}{T_{\rm e}^{1/2}}\,\frac{\omega_{12}}{g_2},
\end{equation}

\noindent where $\omega_{12}$ is the collision strength.  A summary
of line transition data and critical densities is given in
Table~\ref{tab1}.

\begin{table*}
\begin{center}
\caption{Forbidden Line Data$^a$	\label{tab1}}
\begin{tabular}{cccccccccc}
\hline\hline Ion & Transition & $g_1$ & $g_2$ & $\lambda$ & $\beta^b$ &
$\varpi$ & $A_{21}$ & $\Omega^b$ & $n_{\rm c}$ \\
 & & & & $(\mu$m) & & & $(10^{-3})$ s$^{-1}$ & & $(10^4)$ cm$^{-3}$ \\ \hline
Ca{\sc iv}  & $^2$P$_{1/2}\rightarrow ^2$P$_{3/2}$ & 4 & 2 &  3.2 & 0.45  & 3.14 &  545  &  1.06 & 1200 \\
Ne{\sc ii}  & $^2$P$_{1/2}\rightarrow ^2$P$_{3/2}$ & 4 & 2 & 12.8 & 0.113 & 2.24 & 8.55  &  0.30 & 65.0 \\
Ne{\sc iii} & $^3$P$_{1}  \rightarrow ^3$P$_{2}$   & 5 & 3 & 15.6 & 0.092 & 1.83 & 5.97  &  1.65 & 12.6 \\
S{\sc iv}   & $^2$P$_{3/2}\rightarrow ^2$P$_{1/2}$ & 2 & 4 & 10.5 & 0.137 & 0.57 & 7.73  &  6.42 & 5.6 \\
O{\sc iv}   & $^4$P$_{5/2}\rightarrow ^4$P$_{1/2}$ & 2 & 6 & 32.6 & 0.044 & 0.35 & 0.518 &  0.69 & 5.2 \\
Ne{\sc v}   & $^3$P$_{2}  \rightarrow ^3$P$_{1}$   & 3 & 5 & 14.3 & 0.101 & 0.66 & 4.59  &  5.82 & 4.6 \\
S{\sc iii}  & $^3$P$_{2}  \rightarrow ^3$P$_{1}$   & 3 & 5 & 18.7 & 0.077 & 0.65 & 2.07  &  5.81 & 2.1 \\
Si{\sc iii} & $^3$P$_{2}  \rightarrow ^3$P$_{1}$   & 3 & 5 & 38.2 & 0.038 & 0.62 & 0.242 & 10.4  & 0.13 \\ \hline
\end{tabular}

{\small $^a$ Atomic data taken from Pradhan \& Peng 1995.} 
{\small $^b$ Evaluated for $T_{\rm e}=10^4$ K.} 
\end{center}
\end{table*}

Forbidden emission lines in stellar winds are optically thin.  
The above relations are typically combined with the intent
of integrating the emissivity over the wind and solving for the ionic
abundance, derivable from observed
forbidden emission lines and used to constrain gas abundances in
evolved stars to test massive star evolution models.
The primary goal of our paper is quite different.  We seek to
derive line profile {\em shapes} to determine to what extent those
shapes may be used to deduce the properties of binary orbits and
stellar winds in colliding wind systems.  To do so, we first review
the solution for a spherical wind.

\subsection{Spherical Winds}

It is useful to review briefly the line emission for a forbidden
line from a spherically symmetric wind, both to establish notation
and as a reference to use with the non-spherical case.
For a spherical wind with mass-loss rate \mdot\ and radial speed
$v(r)$ for radius $r$, the mass density is

\begin{equation}
\rho_{\rm sph} = \frac{\mdot}{4\pi\,r^2\,v(r)}.
\end{equation}

\noindent The electron number density is $n_{\rm e} = \rho_{\rm
sph} / \mu_{\rm e} m_H$, for $\mu_{\rm e}$ the mean molecular weight
per free electron.

The region of line formation is set roughly by the radius at which
$n_{\rm e} = n_{\rm c}$.  For hot stars $\dot{M} \sim 10^{-10}$ \msunyr\
and larger, with $v_\infty \sim 1000$~\kms\ or more for OB stars (e.g.,
Lamers \& Cassinelli 1999).  To estimate the radius of line formation,
we introduce an electron number density scale factor that depends on these 
basic wind parameters:

\begin{equation}
n_{\rm 0,e} = \frac{\mdot}{4\pi\,R_*^2\,\vinf\,\mu_{\rm e}\,m_H}.
\end{equation}

\noindent Then the critical radius becomes

\begin{equation}
\frac{r_{\rm c}}{R_*} = \sqrt{\frac{n_{\rm 0,e}}{n_{\rm c}}}.
\end{equation}

\noindent With $n_{\rm 0,e}$ of order $10^{13}$ cm$^{-3}$ for a
WR~star, and $n_{\rm c}$ about $10^{5}$ cm$^{-3}$, forbidden line
emission forms in the far wind, around $10^4 R_\ast$.

To determine the emission line shape, we combine
equations~(\ref{eq:emiss}) and (\ref{eq:n2}) to obtain the emissivity:

\begin{equation}
j = j_0 \, \frac{D_{\rm c}\,n_{\rm e}/n_{\rm 0,e}}{1+\varpi+\varpi\,
	D_{\rm c}\,n_{\rm e}/n_{\rm 0,e}}
\end{equation}

\noindent where $j_0$ is a constant that depends on the line of
interest, $D_{\rm c}\ge 1$ is the `clumping factor' that introduces
a dependence on the wind clumping (following Dessart \etal\ 2000).
Note that Ignace \& Brimeyer (2006) did not account for clumping
in their expressions.

The total luminosity of the optically thin line emission
is given by a volume integration involving the emissivity: 

\begin{equation}
L_l = 4\pi\,\int\,j(r)\,D_{\rm c}^{-1}\, r^2\,dr\,d\cos 
	\vartheta\,d\varphi.
	\label{eq:linelum}
\end{equation}

\noindent where $(r, \vartheta, \varphi)$ are spherical coordinates
in the star system, and the inverse of the clumping factor is the
volume filling factor.  Substituting for a normalized radius $x=r/R_*$
and inserting the emissivity function above, one has

\begin{equation}
L_l = \frac{L_0}{4\pi}\,\int\,\frac{dx\,d\cos\vartheta\,d\varphi}
	{1+\varpi+\varpi\,D_{\rm c}^{-1}\,(x/x_{\rm c})^2},
\end{equation}

\noindent where $L_0 = 16\pi^2 j_0 R_\ast^3$, and the line is assumed
to form predominantly over the constant velocity flow where $\rho
\propto r^{-2}$ since $x_{\rm c}\gg 1$ (c.f., Fig.~1 of Ignace \&
Brimeyer 2006).

The angular quantities integrate to $4\pi$, and the radial integral
is of a standard form.  Given that $x_{\rm c} \gg 1$, the total line
luminosity becomes

\begin{equation}
L_l \approx L_0\,\frac{\pi/2}{\sqrt{(1+\varpi)\,\varpi}}\,
	\sqrt{D_{\rm c}}\,x_{\rm c}.
	\label{eq:Ll}
\end{equation}

\noindent Note that a constant clumping factor appears as a square
root coefficient to the critical radius, and so the product acts
essentially as a `transformed' critical radius.  Also note that the
scale constant $L_0$ depends on the mass-loss rate.  The assumption
is that equation~(\ref{eq:Ll}) holds if $\dot{M}$ is {\em already}
a clumping-corrected value.  

This of course is the total volume-integrated emission.  More relevant
to our study is the emission profile shape.  
Optically thin lines that form in a constant expansion
and spherically symmetric wind produce flat-topped emission profiles.
The isovelocity zones are given by

\begin{equation}
v_{\rm z} = - \vinf \, \cos \theta = -\vinf\,\mu,
	\label{eq:vcones}
\end{equation}

\noindent where $\theta$ is the polar angle from the observer axis
$z$.  With $v_{\rm z}$ fixed, the isovelocity zones are cones.
Introducing $w_{\rm z}=v_{\rm z}/\vinf$, and noting that $dw_{\rm
z} = - d\mu$, the profile shape $dL_l/dv_{\rm z} = \lambda^{-1}\,L_\nu$
becomes

\begin{equation}
\frac{dL_l}{dv_{\rm z}} (w_{\rm z}) = \frac{1}{2}\,\frac{L_l}{\vinf}.
\end{equation}

\noindent In the consideration of line shapes from colliding wind systems,
the above constructions will continue to prove useful.

\begin{figure}
\centerline{\epsfig{file=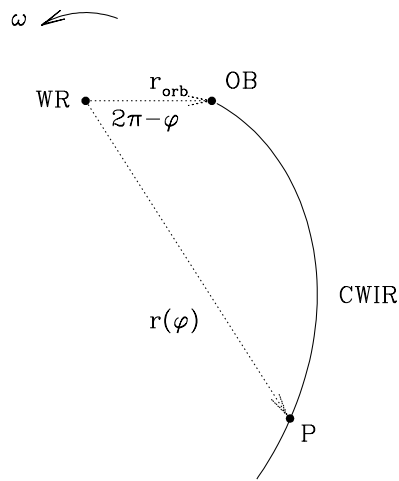,angle=0,width=8cm}}
\caption{
Schematic of the spiral Colliding Wind Interaction Region (CWIR)
in the binary orbital plane.  The WR star is at left and the
OB companion at right.  In this ``snapshot'' the binary separation
is $r_{\rm orb}$.  The orbital motion is counterclockwise as
signified by the angular speed $\omega$, and so the spiral
CWIR winds out in a clockwise fashion.  A point along the CWIR $P$
is indicated as having polar coordinates $r$ and $2\pi-\varphi$.
\label{fig1}}
\end{figure}

\subsection{Colliding Wind Binaries}

Ignace \& Brimeyer (2006) explored the effects of axisymmetric density
distributions for forbidden lines shapes from single star winds.
In that study the key parameters that determined deviations from
flat-top morphologies were the viewing inclination, the latitudinal density
description (whether bipolar or disk-like flow), and the ionization
balance.  The colliding wind systems allow for considerably more
influences, in part because the system is completely non-axisymmetric
and also because there are really two sets of parameters to consider
-- those of the orbit and those of the stellar winds.

In order to explore the range of line shapes and their relevance for
observations, some approximations will be adopted.  We consider WR+OB
systems such that it is the WR wind that dominates the emission of
the forbidden line.  This is reasonable because the mass-loss is much
higher and the abundances are non-solar for the WR component.  For example,
[Ne{\sc iii}] at 15.56 microns is typically prominent across WR subtypes
(Ignace \etal\ 2007), and neon can be enhanced above solar by factors
of a few or much more in a WC wind.  As a result, one major assumption
of our study is that the confined wind of the OB companion as dominated
by the more massive WR wind will have zero emissivity.

As noted before, the forbidden lines form at large radii (i.e., 
compared to the stellar sizes, but not necessarily larger than the orbit
size). We will assume that the WR wind is in constant expansion
for evaluating the line profile shape.

For the binary we choose to work in a frame centered on the WR star
(see Fig.~\ref{fig1}).  The OB companion is taken to follow an
elliptical orbit of semi-major axis $a$ and eccentricity $e$.  The
observer is located on the $z$ axis, and the normal to the orbit
plane is defined to be $z_*$.  The viewing inclination is given by
$\cos i = \hat{z}\cdot \hat{z_*}$.

The WR coordinate system is described by spherical coordinates $(r,
\vartheta, \varphi)$, and Cartesian coordinates $(x_*,y_*, z_*)$.
The observer frame also centered on the WR star has coordinates
$(r, \theta, \alpha)$ and $(x,y,z)$.  The $x$-$z$ and $x_*$-$z_*$
planes are taken as coincident.  Azimuthal angles are measured 
counterclockwise in standard fashion from the $x$ and $x_*$ axes.  The
reference point for the orbit is the periastron location $\varphi_{\rm
p}$.  At any given epoch, the OB star will be located at azimuth
$\varphi_{\rm s}$, which we take to be the orbital phase.

The CWIR will be described in a largely asymptotic form.  We consider
the interaction region to have two primary zones:  a shocked layer for
the OB star wind, and a shocked layer for the WR wind.  The interface
is taken to describe the confinement of the OB wind.  Since we ignore
any line emission from the OB wind, the trace of that interface and
its interior is treated as an emission `cavity'.  Line emission then
arises from two sectors only.  The first is the shocked WR wind.  The second
is an approximately spherical WR~wind that exists outside the shock layer.

The asymptotic shape of the CWIR is that of a spiral, motivated by
observations of `pinwheel' nebula from dust and radio emission in
several WR~binaries (e.g., Monnier \etal\ 2002).  If the OB companion
is at azimuth $\varphi_s$ and orbiting counterclockwise, then the
interaction region spirals outward in a clockwise direction, and
vice versa.  We assume that the bowshock at the OB star has circular
cross-section.  The `memory' of that interaction is moving
radially outward at constant speed. The intersection of the spiral
interaction region with a spherical shell centered on the WR star
is assumed to maintain a constant solid angle of circular cross-section.

The following relations describe the CWIR as related to the binary orbit.
The OB star is located at a radius $r_{\rm orb}$ in the orbital plane
as given by

\begin{equation}
r_{\rm orb} = a\,\frac{1-e^2}{1+e\cos(\varphi-\varphi_{\rm p})}.
\end{equation}

\noindent At this distance the interface between the WR and OB star
winds is taken to have a half-opening angle $\beta$.  The shocked
layer of the WR wind is taken to have a half-opening angle $\beta'$.
Thus the intersection of a spherical shell with the compressed WR
wind layer appears as an annulus of angular width $\Delta \beta =
\beta ' - \beta$.  The characteristic lateral lengths $l$ and $l'$ associated
with these angles are:

\begin{eqnarray}
l & = & r_{\rm orb}\,\tan \beta, \\
l' & = & r_{\rm orb}\,\tan \beta '.
\end{eqnarray}

\noindent Assuming the shock is strong, the immediate post-shock
density will be 4 times that in the normal spherical WR~wind component.
Assuming an adiabatic shock layer with constant post-shock density,
then the annular cross-section of the shocked layer has one quarter the
solid angle of the confined OB~wind, leading to the follow relation that
we adopt for our models:

\begin{equation}
\cos \beta' = \frac{5\cos\beta-1}{4}.
\end{equation}

Note that we treat $\beta$ largely as a free parameter in our models
to explore profile shape effects.  In
fact, its value should be related to the ratio of wind momenta for
the WR and OB star winds.  We define

\begin{equation}
\epsilon = \frac{(\dot{M}\vinf)_{OB}}{(\dot{M}\vinf)_{WR}}.
\end{equation}

\noindent Then the stagnation point between the two stars will be
located at a radius $r_0$ from the WR~star as given by (e.g., 
Shore \& Brown 1988):

\begin{equation}
r_0 = \frac{r_{\rm orb}}{1+\sqrt{\epsilon}}.
	\label{eq:stag}
\end{equation}

\noindent Recently, Gayley, Parsons, \& Owocki (in preparation) have
considered the opening angle associated with a purely adiabatic shock,
which is the case adopted for our work, in contrast to a radiative shock
(see Canto \etal\ 1996; Antokhin, Owocki, \& Brown 2004).  Gayley \etal\
(in preparation) derive the following transcendental relation for the
shock cone opening angle:

\begin{equation}
\cos \beta = \frac{1+\epsilon^2-2\epsilon}{1-\epsilon^2},
	\label{eq:beta}
\end{equation}

\noindent This expression will be used in our application to
WR~147 later in this paper.

Interior to the orbit, the WR wind is assumed spherical.  The bow
shock region in the vicinity of the OB star will be dominated by
hot, X-ray emitting gas (e.g., Parkin \& Pittard 2008).  Consequently,
the requisite ions for contributing to the forbidden lines simply
won't exist there.  As a result, we model the bow shock zone OB
star with cut-offs.  We assume that the confined OB wind (hereafter
referred to as the `cavity') and the compressed WR wind (hereafter
simply the `compressed layer') extends to the stagnation point $r_0$
between the two stas and is too hot to produce forbidden line
emission for the ions of interest.  We further assume that the gas
remains hot some distance downstream of $r_{\rm orb}$.  We adopt
$8r_0$ for this distance, using Cassinelli \etal\ (2008) for
adiabatic shocked wind flow around a blunt object as a crude
guide.  As long as $a\ll r_{\rm c}$ or $a\gg r_{\rm c}$, the details
of the bowshock region will not greatly impact the forbidden line
profile shape.  It is more relevant when $a \sim r_{\rm c}$,
in which case our results can only be taken as illustrative.

It remains then only to trace the center of the cavity with distance
and azimuth in the orbital plane to completely define the spiral
interaction region through the WR wind.  A cross-section of the
spiral pattern is basically a `shadow' of the wind collision that
advances through the WR wind at constant radial expansion.  We
simply need an expression for the star's location around the orbit
with phase to determine an equation of motion for this `shadow'.
Such a relation is derivable from Kepler's laws.

At any given phase, the cross-section center advances radially
according to

\begin{equation}
\dot{r} = \vinf.
\end{equation}
                                                                                
\noindent Conservation of angular momentum ${\cal L}$ provides a
relation for the star location:

\begin{equation}
\dot{\varphi_*} = \frac{{\cal L}}{r_{\rm orb}^2},
\end{equation}

\noindent with the specific angular momentum given by

\begin{equation}
{\cal L} = \sqrt{GMa\,(1-e^2)},
\end{equation}

\noindent for $M$ the summed mass of the two stars.  Defining
an angular velocity $\omega = 2\pi/P$ for period $P$, using 
Kepler's third law, and combining the three preceding equations,
a differential equation for the cross-section center can be
derived:

\begin{equation}
\frac{d\varphi_*}{dr} = -\frac{a^2/r_{\rm orb}^2}{r_{\rm w}}\,
	\sqrt{1-e^2},
\end{equation}

\noindent where a convenient scaling
parameter we call the `wrapping length' $r_{\rm w}$
is introduced:

\begin{equation}
r_{\rm w} = \vinf/\omega.
\end{equation}
                                                                                
\noindent The wrapping length is related to the pitch angle of the
spiral shape.  Since $r_{\rm c}$ is a scale for the emission of the
line, it is natural to characterize models by the ratio $2\pi r_{\rm
w}/r_{\rm c} = P/t_{\rm c}$, where $t_{\rm c} = r_{\rm c}/\vinf$
indicating the number of wrappings per critical radius crossing
time.  If the
wind flow time across the critical radius is much longer than $P$,
the spiral will have many circuits over that scale; but if the
orbital period is long, the spiral is essentially a cone over
the span of a critical radius.

\begin{figure}
\centerline{\epsfig{file=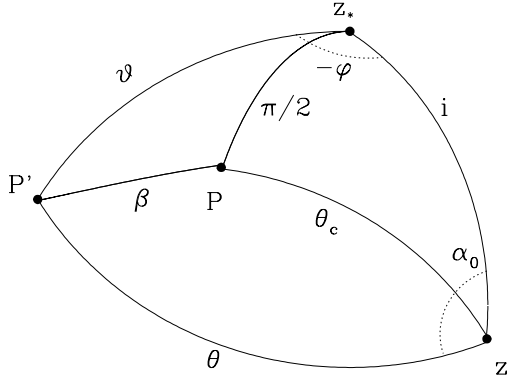,angle=0,width=8cm}}
\caption{
Geometry related to the conical interaction region and the evaluation
of eq.~(\ref{eq:cosbeta}).  The $z$ point is toward the oberver and
the $z_\ast$ is the binary orbital axis.  Point $P$ signifies
the axis of the conical CWIR, and point $P'$ is located on the
shock contact discontinuity.  A similar set of spherical triangles
would apply for a point on the boundary of the compressed layer at
angles $\beta'$ and $\theta'$.
\label{fig2}}
\end{figure}

\subsection{Conical Interaction Regions	\label{sec:conical}}

Before tackling the case of general orbits, it is instructive to
consider the extremely long period binaries for which the bow
shock geometry is asymptotically conical in shape, having no spiral
curvature.  We will refer to this limit as the `conical' 
(or 'linear') bow shock.

In the approximation of negligible wrapping of the spiral, the cavity and
compressed layer are cones centered on the line of centers for the
two stars.  The isovelocity zones are also cones.  An important point
is that we assume the flow in the compressed layer is radial and
has the same speed as the WR terminal speed.  This is a reasonable
approximation for an adiabatic shock and a strong WR wind (see
Tuthill \etal\ 2008).  These isovelocity cones are centered on the
observer's line-of-sight, and so are inclined to that of the
CWIR.  What is interesting is that the intersection
of the interaction region with the isovelocity cones is a fixed
pattern with radius.

Consider a spherical shell.  The cross-section of the isovelocity
zones are rings.  That of the interaction region is a ring also
which, without loss of generality, is assumed to lie in the $x$-$z$
plane.  It is straightforward to find the crossing points between
the two rings in terms of the azimuthal angle $\alpha_0$ for the
observer (see Fig.~\ref{fig2}).  If the cavity center is at
$\theta_{\rm c} = 90^\circ - i$, the implicit solution for $\alpha_0$
is

\begin{eqnarray}
\cos \beta & = & \cos \theta_{\rm c}\cos\theta + \sin\theta_{\rm c}\sin
	\theta \cos \alpha_0 \label{eq:cosbeta} \\
 & = & w_{\rm z} \, w_{\rm c} +\cos \alpha_0\, \sqrt{(1-w^2_{\rm z})(1-w^2_{\rm c})},
\end{eqnarray}

\noindent $w_{\rm z}$ is the observed normalized velocity shift in the
line and $w_{\rm c}$ is the value for the isovelocity zone for the
cavity axis.  The same relation can be used for $\beta'$
and $\alpha_0'$ to determine the crossing points between an isovelocity
ring on the shell and the circular boundary of the compression layer.
The key is that $\alpha_0$ and $\alpha_0'$ are constants with radius.

There are several special cases that are notable.  If viewed pole-on,
intersections only occur for $\theta >90^\circ - \beta'$ and $\theta
<90^\circ + \beta'$.  When viewed edge-on, the axis of the interaction
region coincides with the $z$-axis, in which case there is no
emission for $\theta<\beta$, and emission from the compressed layer
is exclusively from $\beta<\theta<\beta'$.

\begin{figure}
\centerline{\epsfig{file=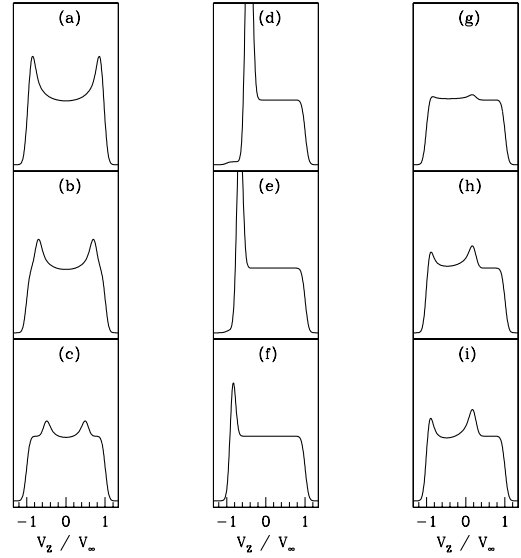,angle=0,width=8cm}}
\caption{
Model line profiles in the `conical bow shock' (i.e., $r_{\rm w}
\gg r_{\rm c}$).  Each profile is normalized to have unit area, and
the horizontal axis is normalized velocity shift.  Table~\ref{tab2}
details the model parameters for each panel.
\label{fig3}}
\end{figure}

\begin{table}
\begin{center}
\caption{Model Runs for Figure~\ref{fig3}	\label{tab2}}
\begin{tabular}{lccc}
\hline\hline Panel & $a/r_{\rm c}$ & $\beta$ & $i$ \\ 
      &               & $(^\circ)$ &$(^\circ)$\\ \hline
(a) & 0.1  & 60 & 0  \\
(b) & 0.1  & 45 & 0  \\
(c) & 0.1  & 30 & 0  \\
(d) & 0.1  & 60 & 90  \\
(e) & 0.1  & 45 & 90  \\
(f) & 0.1  & 30 & 90  \\
(g) & 10   & 40 & 30 \\
(h) & 1    & 40 & 30 \\
(i) & 0.01 & 40 & 30 
\end{tabular}
\end{center}
\end{table}

For emission from the compression region, the density is enhanced
by a constant factor of 4 in our treatment.  In terms of the emission
per unit solid angle, $dL_\nu/d\Omega$, the integration along a
hypothetical radial that lies entirely within the compression layer
will be 8 times greater than for one in a purely spherical wind.
Of course, the compression layer has a solid angle
extent that is 1/4 that of the cavity.  In effect, the
wind collision redistributes
mass in a sector of the WR~flow, and there is a net gain in line
emission above what that sector would have produced if it were
undisturbed.  The wind collision leads in essence to a globally
stuctured clump.  The important point is that one should not expect
CWIRs to conserve flux in the forbidden line as compared to a
spherical wind.

Also, the CWIR does not extend down to
the photosphere of the WR~star, but is significantly offset in radius, as
determined by the location of the stagnation point on the line of
centers between the two stars and the width of the compressed
WR~wind.  This means there is a minimum radius to the CWIR, interior
to which a spherical WR~wind makes a simple flat-top contribution
to the line profile.

Example profiles for the conical bow shock approximation are shown
in Figure~\ref{fig3} for different binary separations relative to
the critical radius $a/r_{\rm c}$, cavity opening angles $\beta$,
and viewing inclinations $i$.  Model parameters for the different
panels are provided in Table~\ref{tab2}.  The principle conclusions
are that:  (a) only a pole-on view to the orbit produces a symmetric
profile, with a double-horned appearance, (b) an edge-on view
produces one that is maximally lopsided, and (c) generally an
asymmetric double-horned profile shape results whose appearance
relates to the viewing perspective and orbital parameters.  Note
that these profiles have been gaussian smoothed to simulate limited
spectral resolution.  Given that typical WR winds have $\vinf \approx
1000-3000$ \kms, smoothing with a gaussian of
HWHM $\delta v/\vinf =0.1$ was
adopted to match roughly the resolution of {\em ISO}'s
SWS06 instrument (de Graauw \etal\ 1996).

\begin{figure}
\centerline{\epsfig{file=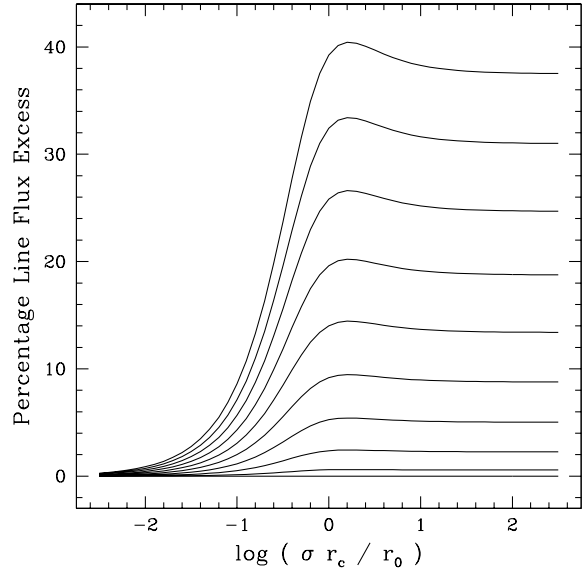,angle=0,width=8cm}}
\caption{
A plot of the integrated {\em excess} line flux for a colliding
wind system as compared to a single spherical wind.  This is plotted
against the ratio of the critical radius associated for the line
relative to the separation of the WR star and OB companion.  An
additional dimensionless multiplicative factor, $\sigma$, accounts
for the wind clumping and atomic constants for the transition
(see eq.~[\ref{eq:sigma}]).
\label{fig4}}
\end{figure}

\begin{figure*}
\centerline{\epsfig{file=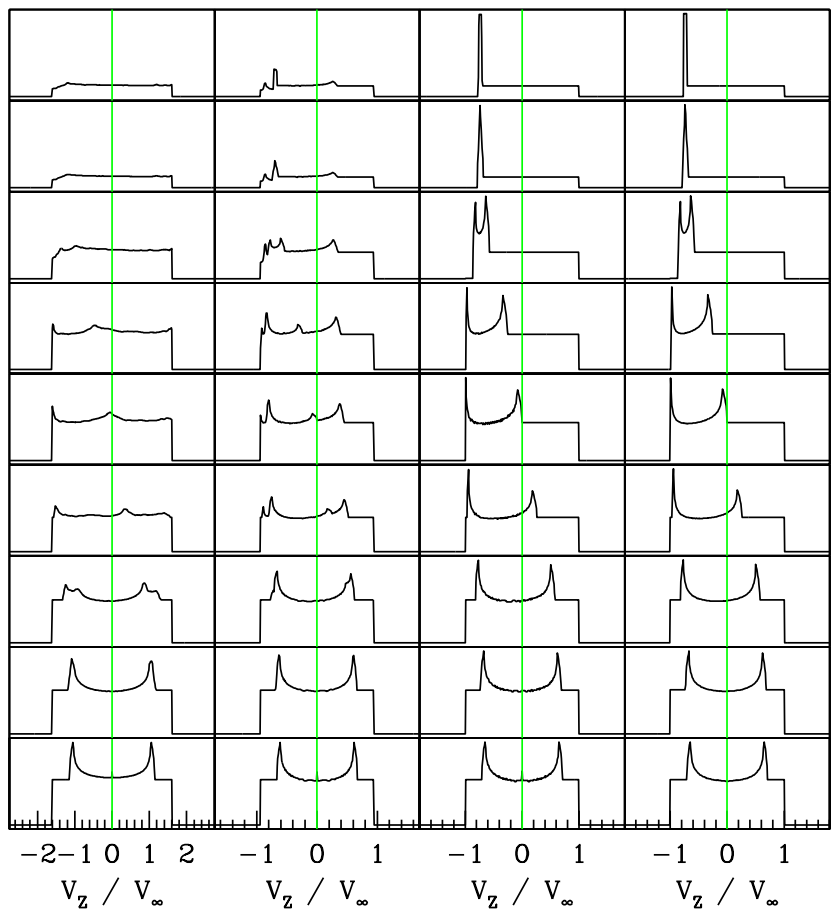,angle=0,width=13cm}}
\caption{
Shown are a series of model calculations to illustrate the range
of profile shapes that result as a function of bow shock parameters,
viewing perspective, and orbital parameters.  The first three columns
are for orbital periods of 1, 100, and 10,000 years, respectively.
The fourth column is for a strictly conical bow shock (essentially
an infinite orbital period, as in Fig.~\ref{fig3}).  From top to
bottom, the viewing inclination is $i=0, 2, 10, 30, 45, 60, 80,
88,$ and 90 degrees.  The orbits are circular ($e=0$).  Notably,
the profile shape is sensitive to the spiral structure, a reflection
of the orbital motion, even for rather long orbital periods of a
century; the effectively conical bow shock limit is not achieved
until a period of about 10 millenia.
\label{fig5}}
\end{figure*}

Finally, all of the examples in Figure~\ref{fig3} have {\em excess}
emission relative to a purely spherical wind by factors of 10-20\%.
It happens that the total line flux as a function of the opening
angle and binary separation is derivable analytically.  There are
four basic zones.  As previously noted for radii $r<r_{\rm orb}$,
the WR wind is spherical and contributes a flat-top contribution
to the profile.  In the cavity sector, there is no contribution.  Then there
is the spherical zone and the compressed layer for $r\ge r_{\rm
orb}$.  Accounting for these zones, the solution for the line
luminosity, relative to a pure spherical wind, is denoted by $\Lambda$
and given by

\begin{eqnarray}
\Lambda & = & 1 + \frac{2}{\pi}\,(1-\cos\beta)\,
	\left[\tan^{-1}\left( \frac{2\sigma\, u_0}{u_{\rm
	c}}\right)\right. \nonumber\\ 
 & & \left. -\frac{5}{8}\, \tan^{-1} \left(\frac{\sigma\, 
	u_0}{u_{\rm c}}\right)\right],
	\label{eq:lineflux}
\end{eqnarray}

\noindent where

\begin{equation}
\sigma = \sqrt{\frac{D_{\rm cl}\,\varpi}{1+\varpi}}.
	\label{eq:sigma}
\end{equation}

\noindent In the limit that $\beta=0$, equation~(\ref{eq:lineflux})
reduces to unity, as it must because there is no CWIR.  For two
identical winds, the opening angle is $\beta= 90^\circ$, and the
maximum line flux for a given value of $\sigma u_0/u_{\rm c} =
\sigma r_{\rm c}/r_0$ becomes:

\begin{equation}
\Lambda = 1 - \frac{5}{4\pi}\,\tan^{-1}\left(\frac{\sigma\, u_0}
	{u_{\rm c}}\right) + \frac{2}{\pi}\,\tan^{-1}\left(\frac{2\sigma\, 
	u_0}{u_{\rm c}}\right).
\end{equation}
                                                                                
\noindent If the binary orbit is exceedingly large, $\Lambda =1$
is again recovered because the CWIR is displaced to a location of
irrelevance with $r_{\rm orb} \gg r_{\rm c}$.  On the other hand,
if $r_{\rm c} \gg r_{\rm orb}$, all the arctangent factors reduce
to $\pi/2$, thus

\begin{equation}
\Lambda = \frac{11-3\cos\beta}{8}.
\end{equation}
                                                                                
\noindent Formally this has a maximum of 1.75.  Figure~\ref{fig4}
shows $(\Lambda-1)$ as a percentage excess of line emission above
the spherical value as a function of $\sigma r_{\rm c}/r_0$.  The
different curves are for different $\beta$ values, from 0 to 90
degrees in 10 degree increments, with the excess being greater for
larger $\beta$.  The location of maximum {\em percentage} excess
can be derived from equation~(\ref{eq:lineflux}) and occurs at
$\sigma r_{\rm c}/r_0 = \sqrt{11}/2$ for all $\beta$.  

These curves are accurate only within the assumptions for the
adopted geometry.  A different set of curves would result for example
if the relation between $\beta '$ and $\beta$ were different.  Also,
for ease of calculation, the bow shock exists for $r\ge r_{\rm rob}$
and not at all for interior radii, which is incorrect because this
neglects the shape of the head of the bow shock.  The exercise does serve to
demonstrate that (a) colliding wind effects will generally not conserve line
flux, (b) line flux excesses at the level of $\sim 10\%$ can be
expected, (c) the excess will depend on the
clumping factor in the large-scale wind, and (d) elemental abundance
determinations for WR winds from forbidden lines formed in colliding
wind systems are not dramatically biased by the CWIR.
However, it may be that changes in ionization between the nominal
spherical wind and the compressed layer could have a larger
systematic influence that would need to be investigated more carefully
in detailed simulations.  We attempt to include such affects in
our consideration of applications to WR~147 and $\gamma$~Vel in Section~3.

\begin{figure*}
\centerline{\epsfig{file=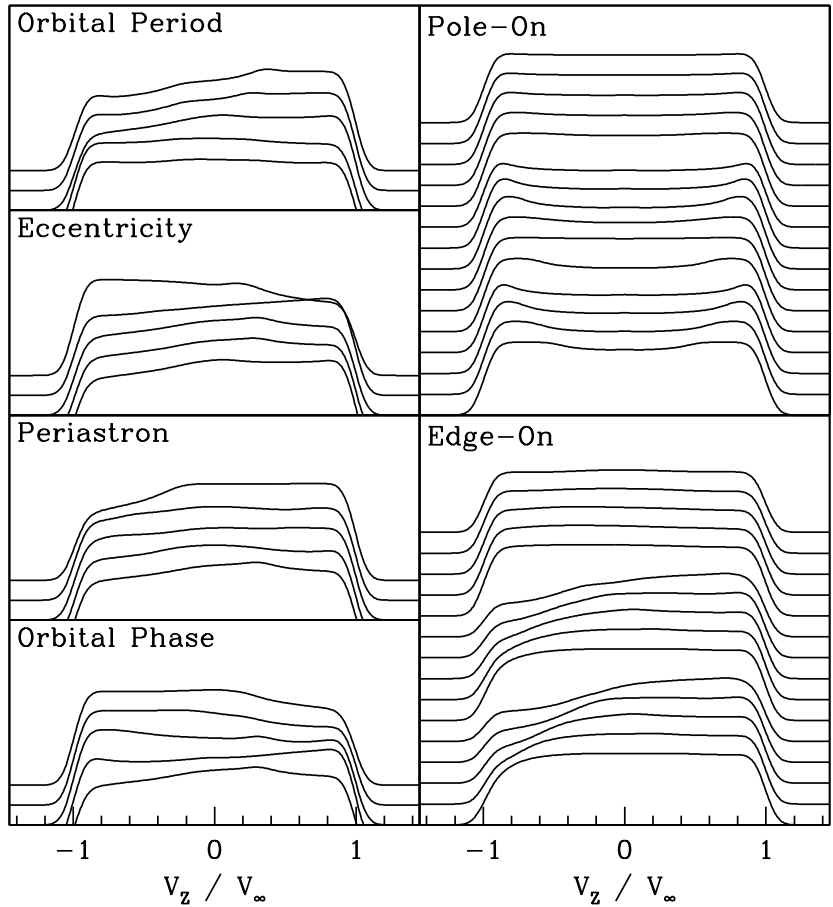,angle=0,width=13cm}}
\caption{
A selection of profile shapes as various individual orbital parameters
are varied.  Labels indicate control parameters.  Each profile is
is plotted against normalized velocity
shift, as in previous figures.  Profiles are vertically offset for
better display of profile shape effects.  Specific
parameter values are described in the text.
\label{fig6}}
\end{figure*}

\subsection{Spiral Interaction Regions}

For spiral-shaped interaction regions when binary orbital effects
are important, issues that influence the profile shape are similar
as for the conical shaped regions of the last section.  However,
there are a few more important factors to bear in mind.  The spiral
is still symmetric about the orbital plane.  That means the line
profile should be symmetric about zero velocity for a pole-on
inclination of the system.  However, when edge-on the spiral shape
now allows the interaction region to transit (eventually) {\em
every} isovelocity zone, thereby tending to dilute the influence
of non-sphericity in the profile shape.  Another important factor
is that the 
direction of orbital revolution with respect
to the observer will be relevant.

One expects the following trends to hold:

\begin{enumerate}

\item Larger opening angles lead to profile structure over a greater
range of observed velocity shifts.

\item Flat-top segments seen toward the line wings places
a limit on $i-\beta'$.

\item The total line luminosity derivation of equation~(\ref{eq:lineflux})
for the conical interaction regions remains relevant for
the spiral case.  With the assumption of constant solid angles, the
proportions of the cavity and compressed layer to a spherical shell
are constant with radius.  This means that a spiral interaction
region is simply a conical one with radius-depdendent phase lags.
However, those curves require a correction for the extension of
the emission cavity down to the stagnation point in regimes where
that portion impacts the line emission.

\item When the wrapping radius is relatively small, the profile
shape tends to deviate {\em less} from a flat-topped appearance,
for most viewing inclinations.

\item The compressed layer leads to line emission being sensitive
to the outer spiral shape even when the wrapping radius is somewhat
large.

\end{enumerate}

The last point is interesting, and is highlighted in Figure~\ref{fig5}
which also provides a test of our code for general orbits against
the semi-analytic case of a conical bow shock.  The figure is organized
in four columns.  From bottom to top, the viewing inclination ranges
from pole-on to edge-on through angles of $i=0, 2, 10, 30, 45, 60,
80, 88,$ and $90^\circ$.  From left to right for the first three
columns, the orbital period is 1 year, 100 years, and 10,000 years.
Profiles in the far right column are for the purely conical CWIR
described in the previous section.  The orbits are taken as circular,
with $r_{\rm orb} = 0.001r_{\rm c}$ so that the CWIR exists essentially
everywhere.  The profile shapes have not been gaussian smoothed.
The bow shock has $\beta = 40^\circ$ and the compressed layer has
$\beta '= 45^\circ$ in each instance.  Note that the vertical scales
(not shown) are the same within a row but vary between rows.

For the longest period case, the profiles in the third
column match those of the strict conical bow shock case in the fourth
column.  What is surprising is that models with $P=100$ years for
which the wrapping radius is larger than the critical radius by two
orders of magnitude deviate from the expectations of the conical bow
shock case.  The reason is that the emission in the compressed zone
is very large.  The factor of 4 increase in density combined with
the density square emissivity dependence in the outer wind along
with the narrowness of the zone leads to relatively `spikey' features
in velocity that appear in the line profile.  Of course with instrumental
smearing, the significance of these features will be substantially
reduced.  Also note that since the geometry is fixed, all the
profiles should have the same flux, which we have confirmed.

It is clear that diverse profiles can result.  Even for the highly
controlled examples of Figure~\ref{fig5}, and with some instrumental
smoothing, profiles can be largely symmetric or strikingly not so;
they may have two peaks, three peaks, or possibly four; and some
profiles can be nearly flat-topped in shape, or have flat-top
segments.

It is difficult to explore the parameter space exhaustively
owing to the large number of free parameters, namely $i$, $\beta$,
$r_{\rm c}/a$, $r_{\rm c}/r_{\rm w}$, $\varphi_{\rm p}$, and
$\varphi_*$ plus the orbit direction, so a limited selection of
illustrative results are shown in Figure~\ref{fig6}.  These have
been convolved with a gaussian of HWHM $\delta v/\vinf= 0.1$.  Each
panel is labeled.  The two longer panels at right are for pole-on
and edge-on viewing inclinations in which the semi-major axis takes
values of $a/r_{\rm c}=0.01, 0.1, 1.0$, five profiles for each case
from bottom to top.  The orbits are circular, and there are five
opening angles $\beta = 30, 40, 50, 60, 70$ degrees.  The orbital
period is 1 year. 
$\gamma$~Vel.

The four panels at left are for more restricted variations in the
parameters as listed.  Generally $P=1$ year, $a/r_{\rm c}=0.1$,
$e=0.1$, $i=60^\circ$, $\beta=40^\circ$, and the periastron and
companion star are both located at $\varphi_{\rm p} = \varphi_{\rm s}
=0$ for these models.
The top panel shows variations in $P$ from 0.01 to 100 years in
factors of ten from bottom top.  For eccentricity, $e=0, 0.05, 0.1,
0.3, 0.8$, again from bottom to top.  The periastron location varies
from 0 to 160 degrees in 40 degree increments; the same variation
is used for the location of the star.  Note that relative to a
spherical wind, the line fluxes are never in excess by more than
23\% for this set of models, with an average excess of about 7\%.

\section{Discussion}

\begin{table}
\begin{center}
\caption{Wind and Binary Parameters for WR~147 	\label{tab3}}
\begin{tabular}{cc}
\hline WR~147$^a$ & WN8+BV \\ \hline
$\dot{M}$ & $25\times 10^{-6}$ $M_\odot$ yr$^{-1}$ \\
$v_\infty$ & 950 km s$^{-1}$ \\
$R_*$ & 21 $R_\odot$ \\
$\mu_{\rm e}$ & 3.1 \\
$D_{\rm cl}$ & 10 \\
$n_{\rm 0,e}$ & $1.2\times 10^{11}$ cm$^{-3}$ \\
$\epsilon$ & 1/90 \\
 & \\
\hline Orbit$^b$ & \\ \hline
$d$ & 630 pc \\
$P$ & 1000 years  \\
$a$ & 870 AU  \\
$e$ & 0  \\
$i$ & $65^\circ$  \\
 & \\
\hline \multicolumn{2}{l}{Interaction Region} \\ \hline
$\beta^c$ & $40^\circ$ \\
$r_{\rm w}$ & 330,000 $R_*$/rad \\
$r_{\rm w}/r_{\rm c}$ (Ca{\sc iv})  & 3300 \\
$r_{\rm w}/r_{\rm c}$ (Ne{\sc iii}) & 330 \\
$r_{\rm w}/r_{\rm c}$ (S{\sc iv})   & 220 \\ \hline
\end{tabular}

{\small $^a$ Morris \etal\ 2000.}
{\small $^b$ Churchwell \etal\ 1992; orbital parameters are
highly uncertain, but given the distance and apparent separation
of the two stars, $a\sim 370/\cos i$ AU.}
{\small $^c$ Using Gayley \etal\ 2009.}
\end{center}
\end{table}

There are currently relatively few WR colliding wind systems that
have adequate data quality for the application of our model to
interpret the CWIR geometries.  {\em ISO} has a spectral resolving
power of about 3000 but observed relatively few WR stars, and 
except for
the fastest WR~winds,
the
spectral resolving power of {\em Spitzer} is too low 
to resolve the features
that our model predicts.  However, {\em ISO} did observe some
colliding wind binaries in its highest resolution mode, and we use
two of them as test case studies: WR~147 and $\gamma$~Vel (WR~11).
The first is a wide binary with an unknown period, although it is
likely on the order of millenia, and the second is a relatively
close binary with an orbital period of a few months.

\begin{figure*}[t]
\centerline{\epsfig{file=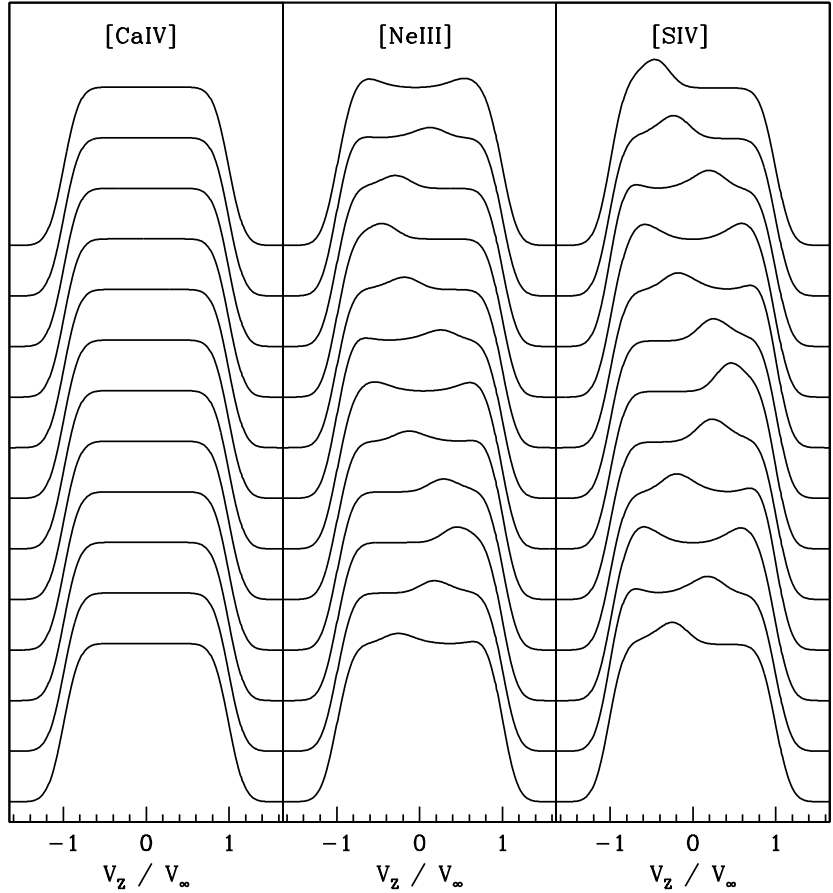,angle=0,width=13cm}}
\caption{
Model line profile shapes as indicated with application to WR~147.
Each profile in a column is for 
a different orbital phase location of the B star companion.  
\label{fig7}}
\end{figure*}

\subsection{Application to WR~147}

Adopted parameters for the WR~147 system are provided in
Table~\ref{tab3}.  The separation of stellar components is at least
360~AU, or about $10^{15}$~cm, based on their projected separation
in the sky (Churchwell \etal\ 1992). The period is estimated to be
at least $P \sim 10^3$ years.  The wrapping radius is thus about
$\gtrsim 10^{17}$~cm.  The wind density scale is around $n_{\rm 0,e}\sim
10^{11}$~cm$^{-3}$.  We assume a statistically likely viewing
inclination of $65^\circ$, in which case the orbital separation
becomes 870~AU.  Having no other knowledge about the orbit, we adopt
$e=0$.  Radio observations by Williams \etal\ (1997) shows that the
bow shock head is quite close to the companion star, 
for which they argue for a BV companion.
The ratio of wind momenta is then $\epsilon \approx 1/90$. 
Using equation~(\ref{eq:beta}) for the bow shock
opening angle for an adiabatic shock, we obtain $\beta \approx 40^\circ$.

Lines of [Ca{\sc iv}], [Ne{\sc iii}], and [S{\sc iv}] were
observed with {\em ISO} (Fig.~8 of Morris \etal\ 2000).  The WR
wind density at the distance of the B star is $n_{\rm e} \approx
10^3$ cm$^{-3}$, well beyond the critical radii values for these
three lines.  Deviations in the emission lines from flat-top should
be mild, consistent with the observations
of Morris \etal  With the sulfur line
forming farthest out, we might expect that some structure would be
observed in it.  In addition, the stellar components are separated
by about $10^4R_\ast$, whereas the winding radius is around 30 (or
more) times that value, thus the CWIR is closely approximated by
the conical bow shock model described in section~\ref{sec:conical}.

Synthetic spectral profiles are displayed in Figure~\ref{fig7}.
Each panel shows model profiles plotted against normalized Doppler
shift.  The different profiles are for different orbital phase
location, sampling a full orbit in $30^\circ$ intervals.  The model
profiles have been convolved with a gaussian of $\Delta v /\vinf =
0.2$ to simulate {\em ISO}'s spectral resolution for the relatively
lower \vinf\ wind of WR~147.  The line of [Ca{\sc iv}] has the
smallest critical radius and shows an essentially flat-top shape.
The neon and sulfur lines show greater variation, as their critical
radii are somewhat closer to the B~star companion.  The observations
of Morris \etal\ (2000) are suggestive of a double-horned feature
in [Ne{\sc iii}] and [S{\sc iv}], and indeed our models manage to
produce such a profile shape in narrow ranges of orbital phase.
However, it is unclear to what extent the profile shapes may be
trusted.  A re-analysis of the [Ca{\sc iv}] line by Ignace \etal\
(2001) suggests an asymmetric line shape.  If real, that could
indicate an elliptical orbit.  A series of model runs with large
eccentricites greater than 0.9 retain double-peaked morphologies
in the Ne and S lines and shows some small deviations from flat-top
in the Ca line for a range of orbital phases.  Although
better S/N data at high spectral resolution is needed to measure
the line shapes more accurately, this preliminary application shows
intriguing potential.

\begin{table}
\begin{center}
\caption{Wind and Binary Parameters for $\gamma$~Vel	\label{tab4}}
\begin{tabular}{cc}
\hline WR~11$^a = \gamma$ Vel & WC8+O7.5 \\ \hline
$\dot{M}$ & $9\times 10^{-6}$ $M_\odot$ yr$^{-1}$ \\
$v_\infty$ & 1550 km s$^{-1}$ \\
$R_*$ & 3.2 $R_\odot$ \\
$\mu_{\rm e}$ & 4.5 \\
$D_{\rm cl}$ & 10 \\
$n_{\rm 0,e}$ & $7.8\times 10^{11}$ cm$^{-3}$ \\
$\epsilon$ & 1/33 \\
 & \\
\hline Orbit$^b$ & \\ \hline
$d$ & 336 pc \\
$P$ & 78.53 days \\
$a$ & 1.2 AU \\
$e$ & 0.334 \\
$i$ & $65.5^\circ$ \\
 & \\
\hline \multicolumn{2}{l}{Interaction Region} \\ \hline
$\varphi_*^c$ & 0.28 \\
$\beta^d$ & $85^\circ$ \\
$r_{\rm w}$ & 750 $R_*$/rad \\
$r_{\rm w}/r_{\rm c}$ (Ca{\sc iv})  & 4.2 \\
$r_{\rm w}/r_{\rm c}$ (Ne{\sc ii})  & 0.68 \\
$r_{\rm w}/r_{\rm c}$ (Ne{\sc iii}) & 0.36 \\
$r_{\rm w}/r_{\rm c}$ (S{\sc iv})   & 0.13  \\ \hline
\end{tabular}

{\small $^a$ de Marco \etal\ 2000}
{\small $^b$ North \etal\ 2007}
{\small $^c$ $T_0=50120.4$ MJD, and {\em ISO} data of $\gamma$ Vel was obtained
on $T=50220.9836$ MJD, hence the O star was at orbital phase 0.28. }
{\small $^d$ Henley, Stevens, \& Pittard 2005}
\end{center}
\end{table}

\begin{figure*}
\centerline{\epsfig{file=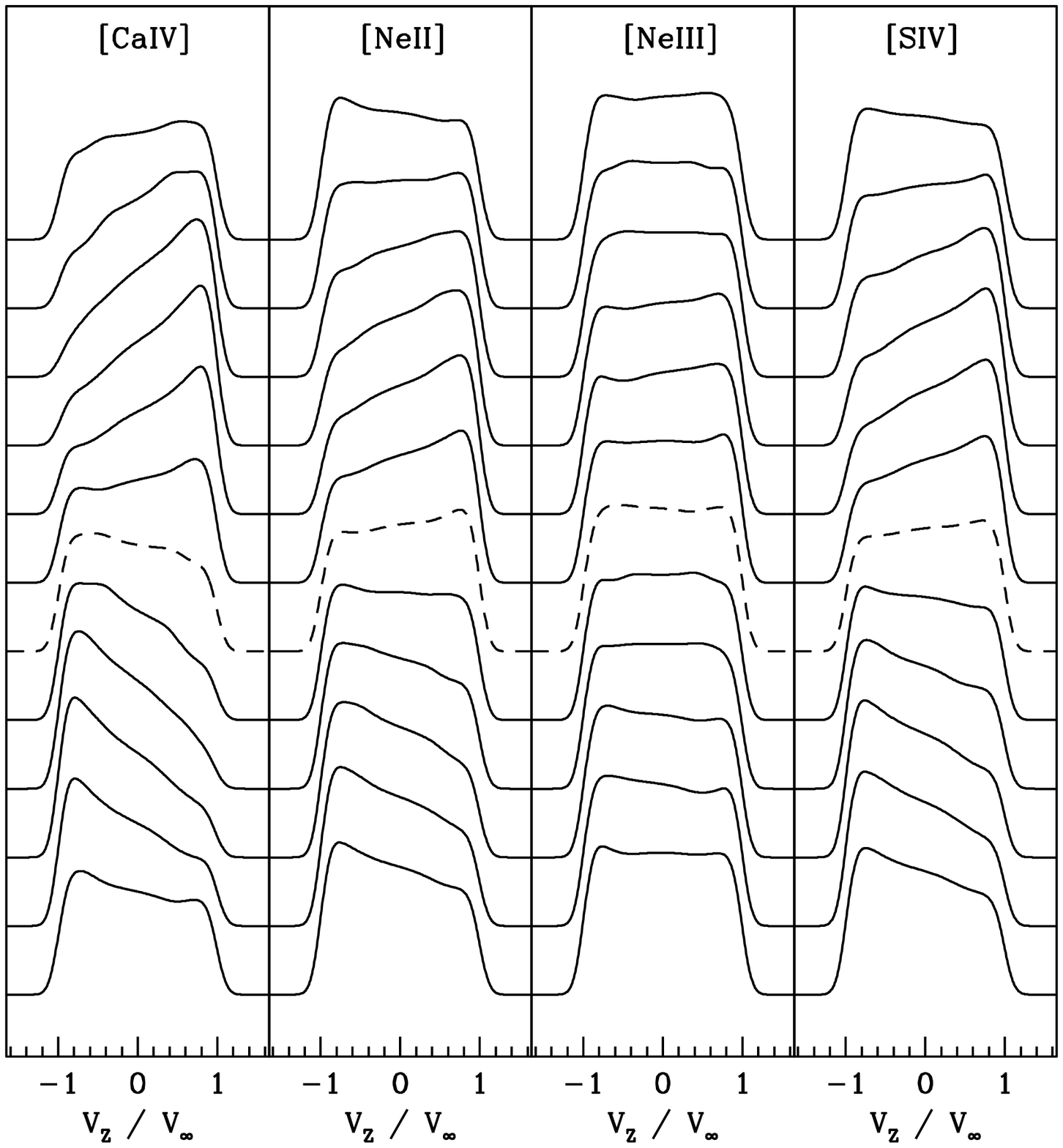,angle=0,width=8cm}\epsfig{file=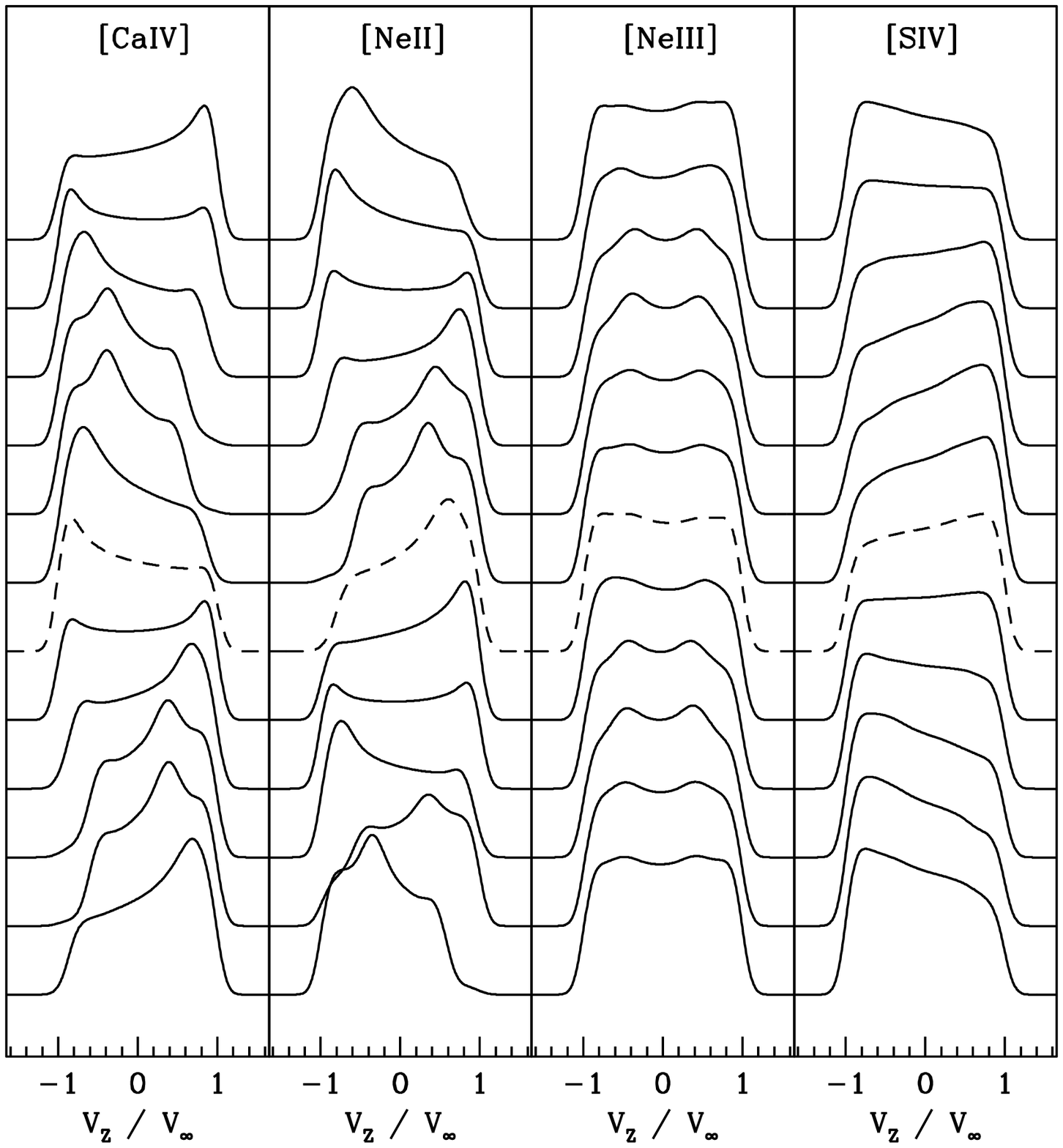,angle=0,width=8cm}}
\caption{
Similar to Fig.~\ref{fig7} except with application to the short
period binary WR~11.  The set of profiles at left assumes constant a constant
ionization factor $Q(r) = Q_0$; the set at right allows for
non-constant $Q(r)$ that best reflects the observed profile shapes
(see text for details).  The dashed line corresponds to an orbital phase
of $150^\circ$, the model closest to $143^\circ$ as determined for
the time that the {\em ISO} data were obtained.
\label{fig8}}
\end{figure*}

\subsection{Application to $\gamma$ Velorum (WR~11)}

Here we apply our synthetic line profile approach to the case of
$\gamma$~Vel consisting of a WC8 star with an O7.5 companion.  Adopted
parameters for the WR wind and the orbit are given in Table~\ref{tab4}.
This colliding wind system has been studied extensively.  The star
and wind parameters come from de Marco \etal\ (2000), and the orbital
parameters come from the interferometric study by North \etal\ (2007).
Note that an independent study by Millour \etal\ (2007) yields mostly
similar parameters.  Four forbidden lines were observed with {\em ISO}:
[Ca{\sc iv}], [Ne{\sc ii}], [Ne{\sc iii}], and [S{\sc iv}] (see Dessart
\etal\ 2000 and Ignace \etal\ 2001).  In contrast to WR~147, the orbital
parameters are quite well-known, and so $\gamma$~Vel constitutes a
significant test case for our modeling.  We note that the adopted opening
angle of $\beta = 85^\circ$ is significantly larger than the value for
an adiabatic shock at $\approx 45^\circ$ (see eq.~[\ref{eq:beta}])
or a radiative shock at $\approx 35^\circ$ (see Canto \etal\ 1996).
We use this larger value based on the X-ray study by Henley \etal\ (2005).

There are a number of new issues that arise in the case of $\gamma$~Vel
that are different from the wide binary WR~147, notably that (a)
the winding radius is comparable to the critical radii of the
observed forbidden lines and (b) the critical radii of most lines
are only somewhat larger than the orbital semi-major (within a
factor of about 10), and for [Ca{\sc iv}], the critical radius is
nearly equal to the separation of the stars at apastron.  
The three primary scale lengths of the problem -- $r_{\rm
w}$, $r_{\rm orb}$, and $r_{\rm c}$ -- are somewhat comparable for
all four lines of interest.  As a result, the single greatest
deficiency of our model becomes potentially important, namely the treatment of
the bow shock head.

Our treatment of the shape of the bow shock interior to the
instantaneous position of the companion star is extremely rough.
We determine the stagnation point $r_0$ based on equation~(\ref{eq:stag}).
The emissivity `cavity' extends inward in a conical fashion from
the companion star to the stagnation point.  This is a gross misrepresentation
of the rounded shape that the bow shock should take.  However, this sector
of the bow shock is the portion where the two winds intersect most nearly
head-on and which produces the observed X-ray emissions (e.g., 
Skinner \etal\ 2001; Henley \etal\ 2005).  This means that the low
ionization states of calcium, neon, and sulfur that contribute to
the IR forbidden line emissions simply will not exist in this sector
at all.  

Figure~\ref{fig8}, left side, shows synthetic emission profiles for
all four observed lines in a format similar to Figure~\ref{fig7}
for WR~147.  The model lines were gaussian smoothed with $\Delta
v/\vinf = 0.14$ to simulate the resolution of {\em ISO} for the
terminal speed of $\gamma$~Vel's wind.  The profiles are only
moderately successful in reproducing the observations.  Dessart
\etal\ found that [S{\sc iv}] was relatively flat-topped in appearance,
whereas the neon lines produce relatively symmetric and triple-peaked
profile shapes.  Ignace \etal\ found that the calcium line was quite
asymmetric, with a strong blueshifted peak.  It is possible
to reproduce the gross profile morphologies for calcium and
sulfur, but nothing like a triple-horned shape results for
the neon lines at any phase.

In order to match the profile shapes of the neon lines, an additional
dependence on radius or direction is required.  Although our treatment
of the bow shock head is very approximate, its shortcomings are not
likely to suppress a three-horned morphology.  Wolf-Rainer Hamann
(private comm.) indicates that in the Potsdam models of
WR~winds, the ionization balance is still not `frozen-in' for WC
models even out to the edge of their computational grid at $200 R_\ast$.
An example of this can be seen in Figure~7 of Gr\"{a}fener, Koesterke,
\& Hamann (2002) where at the lowest wind densities in the model,
ion fractions are still changing with radius.
We thus choose to introduce an additional scaling parameter associated with
varying ionization as given by 

\begin{equation}
Q(r) \propto (n_{\rm e}/n_{\rm 0,e})^p,
\end{equation}

\noindent for some constant exponent $p$.  Given that
$n_{\rm e} \propto r^{-2}$, then $Q \propto r^{-2p}$.

The right side of Figure~\ref{fig8} shows model profiles with
$p = 2$ for Ca$^{3+}$, 1 for Ne$^{2+}$, 2 for Ne$^{1+}$,
and 0 for S$^{3+}$.  Note that changing $p$ to increasingly
positive values tends to move the line formation deeper into
the wind, because the line luminosity scales as

\begin{equation}
L_\nu \propto \int \, j_\nu(r,\vartheta)\, Q(r)\, dV.
\end{equation}

\noindent In addition, the larger density in the compressed layer,
by a factor of 4, tends to enhance recombination, and so for our
models, we allow $p \rightarrow p + 1$ in the shocked layer.
Comparisons to detailed radiative transfer calculations, such as
those of the Potsdam group or CMFGEN (Hillier \& Miller 1999), are
needed to assess the plausibility of the selected $p$ values.
Although motivated by the ionization considerations, $Q(r)$ mainly
acts as an additional weighting function to shift the predominant
region of line formation in radius.

Ultimately, our profile modeling still falls short of
accurately reproducing the observed emission lines.  The [Ca{\sc iv}]
line does become more extreme in its lopsided appearance for some phases.
The [Ne{\sc ii}] line can have a triple-horned appearance at some phases,
which is a good sign; unfortunately, the triple-peaks are never
as equal looking in appearance as observed.  With $Q$ a
constant, the sulfur line continues to display mildly sloped or a mainly
flat-topped appearance for restricted phases.  

There are two main problems with the modeling for $\gamma$~Vel.
First, [Ne{\sc iii}] always shows a more or less double-horned
morphology, never triple-horned as observed.  Second, using the
ephemeris of North \etal\ (2007), we have determined the orbital
phase of $\gamma$~Vel for the {\em ISO} line data.  According to
North \etal, the O star orbits counter-clockwise on the sky, and
so the CWIR spirals outward in a clockwise fashion.  According
to the {\em ISO} archive,
the data were obtained on 1996 May 17 at 23:36:24, or MJD 50220.9836.  
Using the observation of North
\etal\ on MJD 50120.4 and a period of $P=78.53$ days, the O star
is 28\% of an orbit past periastron in time, which corresponds to
an orbital phase of $143^\circ$.  In our simulations this is close
to a phase of $150^\circ$ that is plotted as a dashed line.  Thus,
the second problem is that we cannot simulaneously get all four
profiles with the overall grossly correct morphologies at quite the
same phase.  Note that at a nearby phase of $120^\circ$, the [Ne{\sc
ii}] line does show a triple peak.

\section{Conclusion}

Our study concerns model calculations for forbidden emission line
profile shapes that form in a colliding wind system.  The approach
recognizes that spherically symmetric winds produce optically thin
and flat-topped emission line shapes produced in the constant
expansion flow.  Deviations from that flat-top morphology represents
an opportunity to infer information about the orbital properties
of the binary and the colliding wind bow shock.

Our approach makes a number of significant simplifications to make
the problem tractable:  the OB companion wind makes no contribution
to the line emission, flow in the compressed layer is radial and
with the same speed as the WR wind, the extent of the shocked layer
is only approximate, the geometry of the bow shock head is treated
poorly, the bow shock is assumed axisymmetric about the line of
centers for the two stars, and more input regarding the ionization
balance in the large scale wind is needed to model line shapes
properly.

In light of all of these shortcomings, the models do produce a
remarkable degree of diverse profile morphologies.  Although even
qualitative matches to observed forbidden lines from WR~147 and
$\gamma$~Vel are far from satisfactory, the attempt does appear to
capture a number of trends and shows potential as a tool for deducing
or limiting orbital and wind parameters.  

Although of lower spectral resolution than {\em ISO}, we determined
that the {\em Spitzer} IRS could probably detect modulations of IR
forbidden line shapes with orbital phase.  Unfortunately, the IRS
will not be available during {\em Spitzer's} ``warm'' cycle.  It
is expected that the James Webb Space Telescope will have a
mid-infrared spectrograph\footnote{www.stsci.edu/jwst/instruments/miri}
with a resolving power similar to {\em ISO}.  In addition, the
[Ne{\sc ii}] emission line is observable from ground-based
observatories, as for example in the study of Smith \& Houck (2001).
In particular, their sample included WR~146 which has quite broad
lines owing to a wind terminal speed of nearly 3000 km s$^{-1}$.
So there is the possibility that future observations will provide
new high quality data of these and other systems for which our
diagnostics will be relevant.

\subsubsection*{Acknowledgements}
We wish to thank Ken Gayley for helpful discussions in the early
stages of this project.  We are also grateful a number of useful
suggestions made by an anoymous referee.  This project was funded
by a partnership between the National Science Foundation (NSF
AST-0552798), Research Experiences for Undergraduates (REU), and
the Department of Defense (DoD) ASSURE (Awards to Stimulate and
Support Undergraduate Research Experiences) programs.

\end{document}